\documentclass[conference]{IEEEtran}
\usepackage{epsf,epsfig,subfig, hyperref, amsmath, amssymb, bm,amsbsy, graphicx,xcolor}
\usepackage{tabularx,booktabs}
\usepackage[noadjust]{cite}
\newcolumntype{C}{>{\centering\arraybackslash}X}
\usepackage[english]{babel}
\usepackage{balance}
\usepackage{dblfloatfix}
\usepackage{lipsum}
\usepackage{flushend}
\graphicspath{{./Images/EPS/}}

\usepackage{tikz}
\usetikzlibrary{shapes,arrows}
\usetikzlibrary{calc}
\usetikzlibrary{positioning}
\renewcommand\IEEEkeywordsname{Keywords}
\begin{document}
\title{Latency Analysis of Vehicle-to-Pedestrian C-V2X Communications at Urban Street Intersections}
\author{Kuldeep S. Gill\\
\normalsize Department of Electrical and Computer Engineering, Worcester Polytechnic Institute, Worcester, MA\\
\normalsize Email: ksgill@wpi.edu}

\maketitle
\begin{abstract}
Cellular Vehicle-to-Everything (C-V2X) technology promises to provide ultra-reliable low latency communication (URLLC) framework for connected vehicles. Connected vehicles can help us improve traffic safety, congestion and reduce fatal accidents. C-V2X promises new levels of connectivity and intelligence providing numerous features like infotainment, always-on telematics, real-time navigation, etc. with heterogeneous network architecture. C-V2X leverages existing cellular architecture and is on the way to replace DSRC/WAVE for vehicular communication. There are still some challenges which needs to be tackled before C-V2X can be deployed for efficient on-road V2X. In this paper, we look at end-to-end latency for Vehicle-to-Pedestrian (V2P) which demands ultra low latencies with reliable packet delivery ratio. To evaluate the LTE-V2P performance, we used network simulator for an urban intersection scenario where vulnerable road-side users (VRUs) are trying to cross the street and they communicate with moving vehicles. We evaluated the end-to-end latency and throughput for the given scenario and concluded that with the existing network architecture the latency is high. By utilizing multi-edge access computing (MEC) servers latency can be reduce drastically and hence can be made feasible for cellular V2P communication.
\end{abstract}
\vspace{2pt}
%%%%%%%%%%%%%%%%%%%%%%%%%%%%%%%%%%%%%%%%%%%%%%%%%%%%%%%%%%%%%%%%%%%%%%%%%%%%%%%%%%%%%%%%%%%%%

%%%%%%%%%%%%%%%%%%%%%%%%%%%%%%%%%%%%%%%%%%%%%%%%%%%%%%%%%%%%%%%%%%%%%%%%%%%%%%%%
\section{Introduction}
\label{sec:intro}
Vehicles are gradually moving towards context awareness system where they are aware of their environment . The current vehicular systems heavily relies on direct line-of-sight for context awareness. Connected vehicles can exchange driving environmental information via Basic Safety Messages (BSM) within a transmission range of 500 meters~\cite{wang2017performance}. The BSM can carry information regarding current position, speed of the vehicle, direction, etc. and can provide critical support for vehicular communication~\cite{jiang2008ieee}.

%%%%%%%%%%%%%%Generic-Introduction%%%%%%%%%%%%%%%%%%%%%%%
Connecting vehicles by leveraging wireless communication and networking solutions has been studied intensively in the past, especially with respect to Vehicle-to-Vehicle (V2V) and Vehicle-to-Infrastructure~\cite{eichler2007performance}. IEEE 802.11p DSRC/WAVE~\cite{li2010overview} standard was the first framework designed to meet demands of Vehicular Network (VANET). Although 802.11p was a good starting point but there are some obvious shortcomings of using the standard such as low reliability, hidden node problem, unbounded delay and sporadic Vehicle-to-Infrastructure (V2I) connectivity~\cite{8469236, bilstrup2009does}. 3GPP Long Term Evolution (LTE) has been proposed to mitigate some of these drawbacks of 802.11p with LTE Release 15 where the main focus is on low latency for vehicular communication~\cite{8403963}. C-V2X which uses LTE infrastructure for vehicular communication is now seen as a mainstream reality for future connected cars.

%%%%%% current state-of-the-art%%%%%%%%%%%%%%%%%%%%%
The primary driving force for C-V2X is the LTE backhaul network which can directly be ported for vehicular communication without spending billions on setting up the entirely new infrastructure. Studies are now being conducted to evaluate the feasibility of LTE for vehicular networks~\cite{9128871}. In~\cite{ge2016vehicular}, authors proposed cooperative small-cell network architecture for dense urban canyons. Due to considerable speed of vehicles the topology of VANET becomes highly time-varying which can cause recurrent link intermediate. For such scenarios, cooperative transmission is recommended as an encouraging solution for connected vehicles. The authors in~\cite{8442825} leveraged Multi-access Edge Computing (MEC) technology for cellular network architecture to meet the stringent latency requirement for Vehicle-to-Pedestrian (V2P) communication. By utilizing an overlaid MEC deployment, the authors suggest that achieving very low latencies is possible due to its close proximity to end systems. They numerically evaluated the end-to-end latency for communication links between Vulnerable Road Users (VRUs) and vehicles with and without MEC deployment. With the proposed deployment of overlaid MEC the authors gained an average 80\% reduction in latency in comparison to the existing network architecture.

%%%%%%%%%%%%%%%Problem with the current state-of-the-art%%%%%%%%%%%%%%%%%%%%%%%
The numerical simulation can give a good insight into the performance metrics for V2P communication. However, it is crucial to perform system-level simulation to get more accurate analysis of latency values for the scenario. In this paper, we propose a system-level simulation for communication links between VRUs and vehicles and validate the results from numerical simulation. Latency is a very important parameter for V2P communication but an accurate estimate of Packet Delivery Ratio (PDR) is also needed to benchmark the performance. Lower values of PDR will force frequent Hybrid Automatic Repeat Request (HARQ) which indirectly will lead to higher overall latencies. Network Simulator (NS-3)~\cite{ns} is used for system-level simulation with LTE and flow monitor modules. Traffic flow for the NS-3 is designed in Simulation of Urban Mobility (SUMO)~\cite{sumo} where the trace for an urban intersection is utilized for the network. NS-3 provides a realistic modeling for network layer but does not have enough support for accurate PHY modeling. To accurately model the PHY layer, fading trace for Extended Pedestrian A (EPA) and Extended Vehicular A (EVA) are created in MATLAB, and then imported to NS-3.

\begin{figure*}
\includegraphics[width=\textwidth]{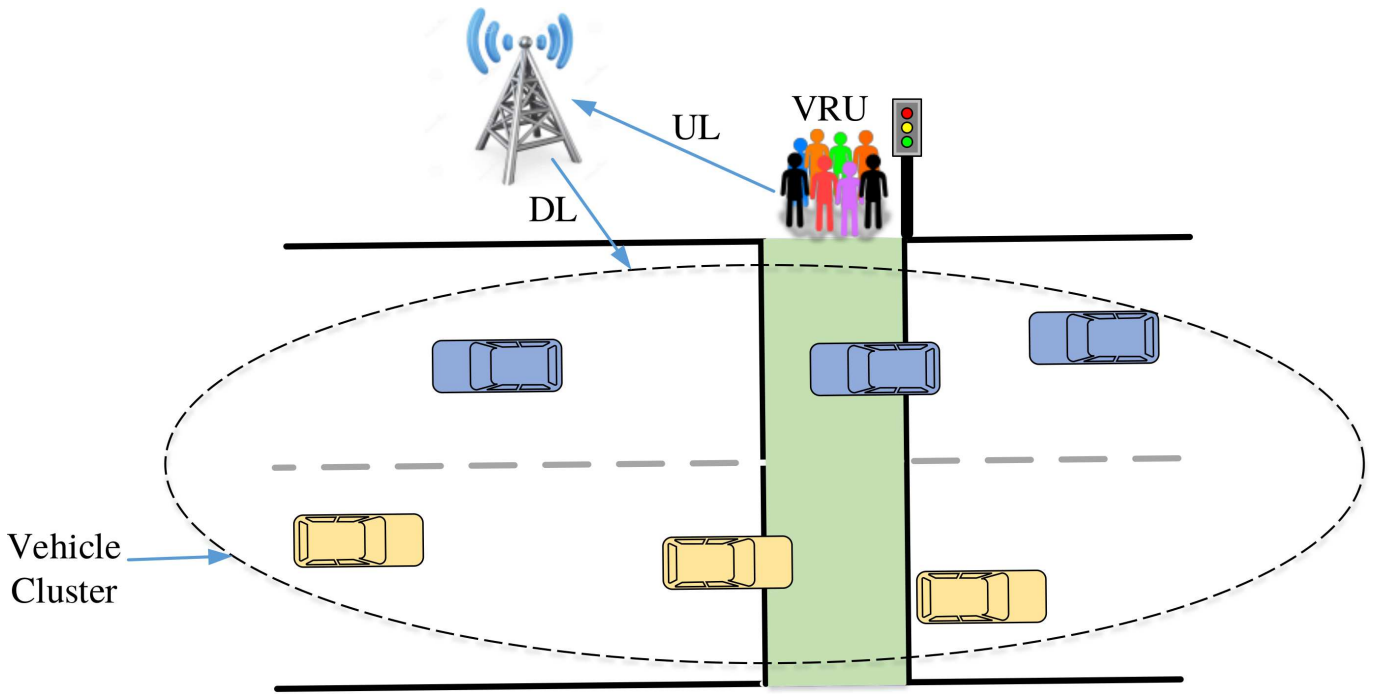}
\caption{Vehicle-to-Pedestrian in an urban intersection where roadside users are trying to cross the street.}
\label{fig:introdiag}
\end{figure*}

%%%%%%%%%%%%%%%%Propose Solution%%%%%%%%%%%%%%%%%%%%%%%%%%%%%%%%%%%%%%%%

This paper is organized as follows: In Section~\ref{sec:trafficflow}, we describe the urban intersection scenario for evaluating end-to-end latency for V2P communication. In Section~\ref{sec:simsetup}, we go through the simulation setup for our scenario where we use SUMO to generate the traffic flow trace and NS-3 for network simulation. We present our simulation results in Section~\ref{sec:results} and finally we conclude the paper with Section~\ref{sec:conclusion} where we discuss the future work for the research.
%%%%%%%%%%%%%%%%Paper Organization%%%%%%%%%%%%%%%%%%%%%%%%%%

%%%%%%%%%%%%%%%%%%%%%%%%%%%%%%%%%%%%%%%%%%%%%%%%%%%%%%%%%%%%%%%%%%%%%%%%%%%%%%%%%%%%%%%%%%%%%%%%%%%%%%%%%%%%%%%%%%%%%
\section{Traffic Flow Generation for Urban Intersection}
\label{sec:trafficflow}
SUMO is a very popular tool for traffic flow generation in the VANET community. It provides a very realistic dynamics of vehicular mobility by utilizing accurate mathematical models. Vehicles brake when they approach a traffic signal, accelerate, decelerate depending on the lane they are driving. With openstreetmap~\cite{osm} API a city-wide traffic analysis can be performed by importing all the static objects like buildings, trees, parks, etc. into SUMO. The abstraction provided by SUMO leave all the room for designing routing algorithms and link analysis with high accuracy. Since the shadow fading and small-scale fluctuations caused by tree foliages, buildings play an important role in the packet delivery ratio, latency, using SUMO can help us get a very accurate estimation of these metrics.

Figure~\ref{fig:introdiag} describes the scenario selected for end-to-end latency evaluation for V2P communication link. In the given scenario, VRUs are trying to cross the street while vehicles are moving along the freeway. The particular scenario is prone to high number of fatal accidents as blind, old VRUs can find it difficult to cross the street and can get hit by on-coming traffic. Latency for this scenario has to be specially low to avoid any delay in information exchange to avoid any accidents. In LTE architecture for any inter-node communication traffic has to be routed via base-station. A node first needs to perform uplink to send the packets to eNodeB, from there the packets are transmitted to receiving node via downlink. Since VRUs and vehicles are not directly exchanging information via direct links, this can lead to large amount of latencies. And if the latency evaluated is high, design changes are required in the C-V2X to accommodate such latency-critical scenarios.

\begin{figure}[!ht]
\includegraphics[width=0.495\textwidth]{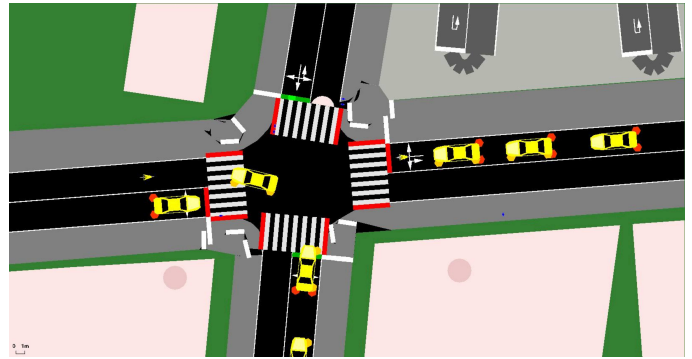}
\caption{Traffic flow trace designed in SUMO for an urban intersection comprising vehicles, pedestrians and cyclists.}
\label{fig:sumo}
\end{figure}

The urban intersection scenario was generated using SUMO and openstreetmap API, where vehicles and pedestrians were generated randomly. They enter and exit the simulation system at random times~\cite{8347136}. Figure~\ref{fig:sumo} shows the traffic generation model created in SUMO, vehicle density was changed gradually to see the effect on packet delivery ratio and average latency. The speed of vehicles was chose from a uniform random variable $\mathbf{U} \thicksim (70,110)$ km/hr and pedestrians velocity was fixed at 3 km/hr. 
%%%%%%%%%%%%%%%%%%%%%%%%%%%%%%%%%%%%%%%%%%%%%%%%%%%%%%%%%%%%%%%%%%%%%%%%%%%%%%%%

\section{Simulation Setup Using SUMO and NS-3}
\label{sec:simsetup}

\subsection{SUMO and NS-3}
Network simulation was performed using NS-3 with LTE and FlowMonitor modules for designing Evolved Universal Mobile Telecommunication System Terrestrial Radio Access Network (E-UTRAN ) architecture, and computing performance metrics respectively. The urban intersection traffic flow model trace was imported into NS-3 using Ns2MobilityHelper class from ns-3 module library. In NS-3 simulator,  LTE Uu protocol stack is implemented on top of all the vehicles and pedestrians where higher layers are encapsulated into transport layer. LTE Uu protocol stack has five five layers namely: Packet Data Convergence Protocol (PDCP), Radio Link Control (RLC), Medium Access Control (MAC), Physical Layer (PHY) and Radio Resource Control (RRC). RRC layer only carries control and signaling information between User Equipment (UE) and eNodeB whereas other layers take care of ciphering, segmentation, HARQ, etc. Non-Access Stratum (NAS) layer has a logical link to Mobility Management Entity (MME) and is responsible for authentication for new UEs joining the E-UTRAN. Figure~\ref{fig:protocol} describes the protocol stack and different roles played by each layer and the direction of datum among the layers.

\begin{table}[t!]
\centering
\caption{Simulation parameters used for evaluating end-to-end latency in V2P technology.} 
\begin{tabular}{| c | c | c |}
\hline
\textbf{Entity} & \textbf{Parameter} & \textbf{Value}  \\
\hline
         & Speed (km/hr) & $\thicksim~\mathbf{U}~(70,110)$ \\
Vehicles & Inter-vehicle distance (m) & sumo traffic model \\
		 & $\lambda$ (Vehicle Density)  & (0.01, 0.09) \\
		 & Cluster size               & Broadcast \\
\hline
         & Number of VRUs (N) & 80\\
VRU      & Tx power (dBm)     & 23\\
		 & Packet size        & 10 kbits\\
\hline
	     & Tx power (dBm)     & 46\\
eNodeB   & Bandwidth (MHz)    & 10 \\
         & Center Frequency (GHz) & 5.9\\
\hline
General  & Channel Model      & Extended Vehicular A Model\\
\hline
\end{tabular}
\label{tab:1}
\end{table} 

\begin{figure}[!ht]
\includegraphics[width=0.4\textwidth]{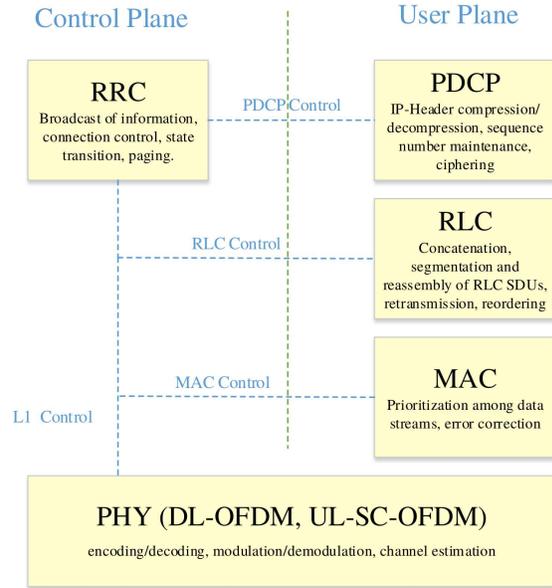}
\caption{LTE Uu protocol stack implemented on eNodeB and vehicles, pedestrian for V2P communication link.}
\label{fig:protocol}
\end{figure}

\begin{figure}[!ht]
\includegraphics[width=0.495\textwidth]{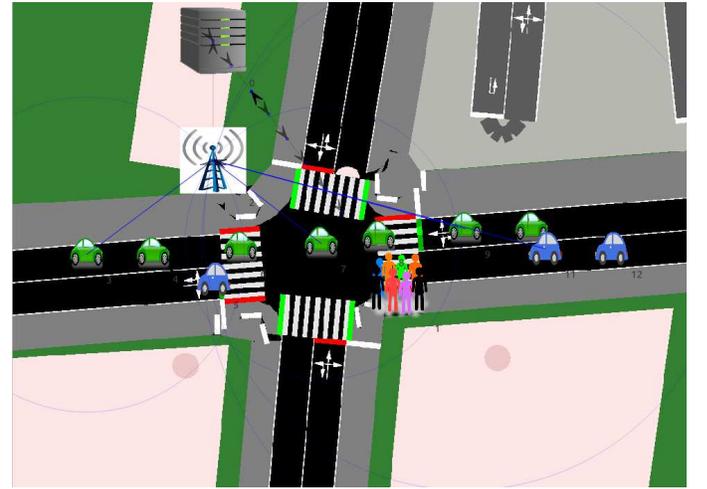}
\caption{Network animation of the LTE V2P architecture for the urban intersection scenario where PGW serves as a backhaul network.}
\label{fig:protocol}
\end{figure}

\begin{figure}[!ht]
\includegraphics[width=0.495\textwidth]{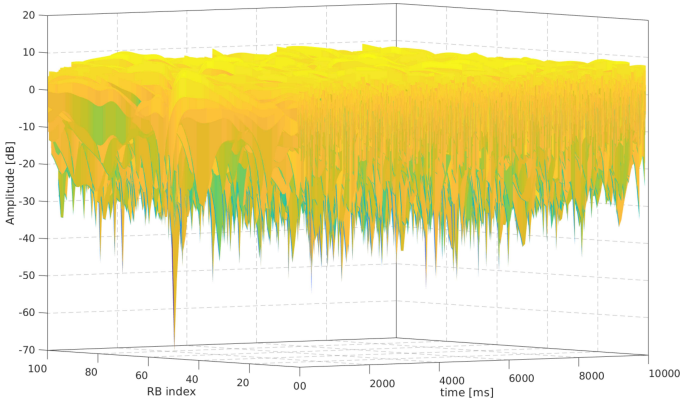}
\caption{Physical layer fading trace for extended vehicular model for velocity $v~=~80~kmph$.}
\label{fig:eva}
\end{figure}

The simulator at the radio level has the granularity of the Resource Block (RB) which is the fundamental block of LTE for resource allocation. NS-3 provides a realistic framework for simulating higher layers but does not provide accurate model for Physical Layer simulation. For modeling PHY layer, fading trace is generated using MATLAB~\cite{matlab} for Extended Vehicular A (EVA) for fixed velocity of 80 kmph, although in the simulation we are assigning velocities from uniform random distribution. High mobility leads to Doppler shift and high time-varying channel which can be model by creating one trace applicable for all the vehicles as the difference would not change the channel propagation conditions drastically. Amplitude variation with respect to RB and propagation time is shown for the EVA propagation model in Figure~\ref{fig:eva}. Extended Pedestrian A (EPA) model is used for simulating the propagation scenario of pedestrians with fixed velocity of 3 kmph. Table~\ref{tab:1} shows the different parameter settings used for vehicles and VRUs in the simulation setup.

\subsection{Latency Modeling}
In this paper, our main motivation is to evaluate end-to-end latency for V2P communication scenario as the high latency value can lead to fatal accidents in these scenarios. We have built the network scenario in SUMO and NS-3 and here we will describe various latency components involved in the packet transmission from VRUs to on-road vehicles. For a conventional cellular network, VRU to eNodeB latency is given by Eq. (\ref{eq:1}):

\begin{equation}
T_{one-way} = T_{UL}+T_{BH}+T_{TN}+T_{CN}+T_{Exc}
\label{eq:1}
\end{equation}

where $T_{UL}$ is the uplink transmission latency from VRU to vehicle, $T_{BH}$ is the backhaul network latency, $T_{TN}$, $T_{CN}$ are transport and core network latencies and $T_{Exc}$ is the processing latency at Evolved Packet Core (EPC) network. Similarly we have the latency incurred due to downlink transmission where the packets are transmitted from eNodeB to vehicles. The resulting E2E latency can be described by Eq. (\ref{eq:2}):

\begin{equation}
T_{E2E} = T_{UL}+2\underbrace{T_{BH}+T_{TN}+T_{CN}}_{Network~Latency}+T_{Exc}+T_{DL}
\label{eq:2}
\end{equation}
where $T_{DL}$ represents the downlink transmission latency. Network simulator internally takes care of the latency modeling and we can create realistic network architecture by using well-defined network attributes. For a MEC-assisted network, we do not need to connect the Packet Gateway (PGW) to the core network and can do all the packet processing on PGW thus reducing the network latency. In Fifth Generation New Radio (5G-NR)~\cite{lien20175g} interface they have already proposed a network architecture where all the packet processing is done on the edge server and that has lead to huge gain in terms of latency. For achieving reliable connectivity in V2P communications where latency is a major hurdle, 5G-NR network architecture can help C-V2X for achieving low latencies.

\section{Simulation Results}
\label{sec:results}

Latency evaluation is conducted for the urban intersection scenario for vehicle to VRU communication link by varying the number of vehicles on a lane per km. We start by 10 vehicles and gradually increase the number of vehicles to 90. We designed a broadcast communication where VRUs at random times, transmit a packet to all the vehicles in the communication range. Since these are Cooperative Awareness Messages (CAMs) it can be broadcasted to all the vehicles near the vicinity containing the information regarding the VRUs position, velocity, etc. where vehicles can implement appropriate actions based on the information.

\begin{figure}[!ht]
\includegraphics[width=0.495\textwidth]{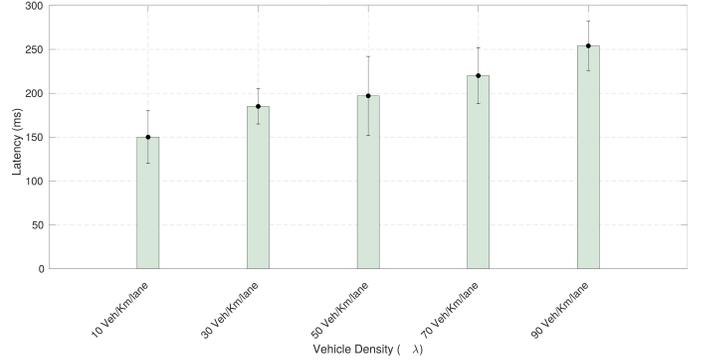}
\caption{End-to-end latency for V2P communication where VRUs randomly generates packets and transmit to the vehicles via eNodeB.}
\label{fig:latency}
\end{figure}

Figure~\ref{fig:latency} shows the end-to-end latency for varying vehicular density, as it is expected latency increases with number of vehicles. The simulation is ran 100 times for each of the scenario and the respective error bars are also plotted on the figure. The end-to-end latency is computed for conventional architecture and as we can see the latency values are high for a practical V2P communication technology. For next part, core network for V2P communication is modified to simulate the multi-edge access computing server. The PGW is just used to simulate MEC as PGW is required to simulate IP-based forwarding in LTE. As we can see in Figure~\ref{fig:latency} latency drastically reduces if we do all the processing on the cloud server rather than backhaul network. We see a huge latency gain compared to the conventional LTE architecure. 

\begin{figure}[!ht]
\includegraphics[width=0.495\textwidth]{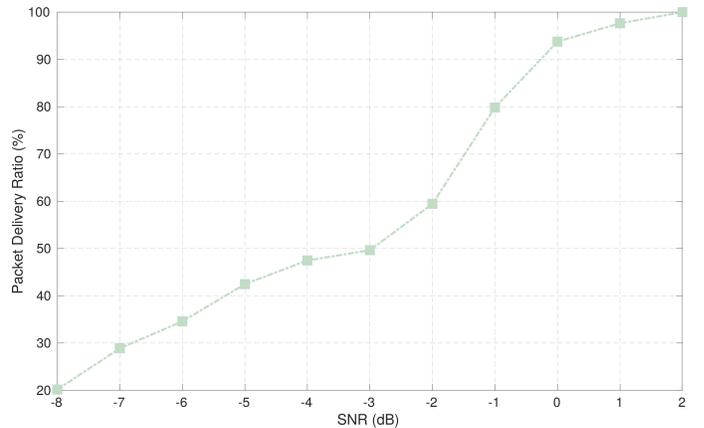}
\caption{Packet delivery ratio for physical downlink shared channel with all resource blocks modulated with 16QAM and 1/2 rate Turbo encoder.}
\label{fig:pdr}
\end{figure}

We also evaluated the packet delivery ratio for Physical Downlink Shared Channel (PDSCH) for conventional LTE architecture by varying the transmit power of eNodeB. The resource block consists of $12$ subcarriers across frequency domain and seven OFDM symbols all modulated with $16$ Quadrature Amplitude Modulation ($16$-QAM). Figure~\ref{fig:pdr} shows the plot for PDR verus signal-to-noise ratio (SNR) and at -$8$ dB the packet delivery ratio is just $20$\% which means that only $20$ out of $100$ packets are delivered successfully. The transmit power of the eNodeB was fixed to  $46$ dBm whereas the transmit power of VRU was set at $23$ dBm. The packets drop were registered for each vehicle and VRU during both uplink and downlink packet transmissions.

\begin{figure}[!ht]
\includegraphics[width=0.495\textwidth]{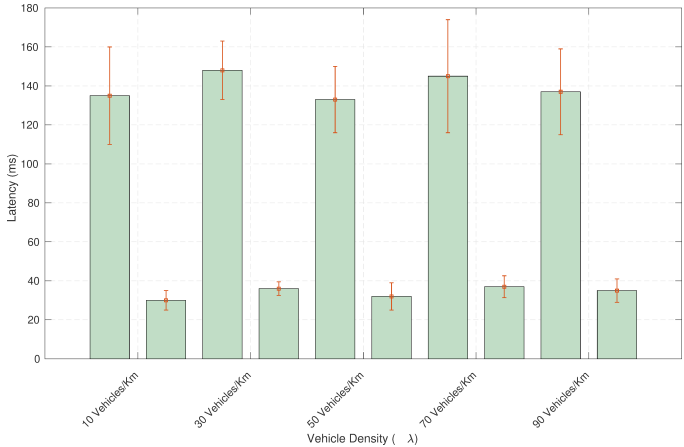}
\caption{End-to-end latency for V2P communication for vehicles near to VRUs and with MEC-assisted procesing of transmission packets.}
\label{fig:latency_small}
\end{figure}

Finally, we also evaluated end-to-end latency using vehicles near to the VRU cluster to get a better comparative performance analysis. We also computed the latency with the MEC-assisted network and we can see there is a huge latency gain if we do not use core network for processing of packets. Figure~\ref{fig:latency_small} shows the latency evaluation with respect to vehicle density. The increase in vehicle does not affect the latency computation as we only select the vehicles which are closest to the VRU cluster.

\section{Comparative Analysis}
In~\cite{8442825} the authors evaluated the end-to-end latency for V2P communication technology using numerical simulation for a freeway segment under cellular coverage. The authors assumed matern hardcore point process for distribution of vehicles across the freeway with uniform distribution for velocities between $70$ and $140$ km/h. The repulsion parameter for hard-core point process was selected as $10$ meters. The VRUs are assumed to be standing in the middle of the freeway trying to cross the freeway, different VRU densities are considered from $50$ to $130$. The authors also assumed that the packets from the VRUs are generated with random timing offsets and their numerical latency was also impacted due to frequent packet collisions among different VRUs. For the core and transport network, authors assumed realistic latency values from a uniform distribution between $15$ and $35$ ms. The authors fixed the value of eNodeB transmit power at $46$ dBm and use $23$ dBm transmit power for VRUs. All vehicles and VRUs were assumed to be serviced by the eNodeB and the authors chose \texttt{WINNER+project}~\cite{meinila2009winner} for their path-loss model. The results gave a good insight on the latency values for freeway intersection scenario. But to get the realistic latency values, a more accurate system level simulation is needed.

In this paper, we build on the author's work and conduct a system level simulation to evaluate the end-to-end latency for V2P communication. We used sumo to generate the traffic flow model and then imported the model into ns-3 and network layer simulation. Since ns-3 does not have a support for PHY implementation, we used MATLAB to create the fading trace and used it with ns-3. Our results deviate from the numerical simulation as we did not perform the multicast to select group of vehicles. In the paper, the authors were multicasting to a vehicle cluster wherein increasing the number of vehicles decrease their latency as the probability of the cluster being closer to VRUs increases. In this paper, we are doing broadcast where the eNodeB will broadcast any packet received from the VRUs to all the vehicles in the transmission range. Hence, we see in Figure~\ref{fig:latency} that our latency actually increases as we increase the vehicular density. To do a performance comparison with the paper, we chose the nearest five vehicles and re-evaluated the latency again to see how much it deviated from the paper.

\section{Summary and Future Research Directions}
\label{sec:conclusion}
In this paper, we evaluated the end-to-end latency for V2P communication using the conventional cellular architecture and then we compared it with MEC-assisted network where we saw a huge latency gain of 75.64\%. We found that using conventional cellular architecture, it is not going to be feasible to facilitate V2P technology as the latencies will be too high to support the communication. Hence, MEC-assisted architecture needs to deployed for efficient and delay-constraint V2P communication. We implemented the simulation test-bed using SUMO, a traffic flow generation tool and then imported into a network simulator NS-3. We compared our results with the paper~\cite{8442825} and found out that our latency did not vary much with the increasing density whereas in their case the value saw a decline in the latency. The reason for a such a difference in latency was because they chose a multi-cast simulation scenario where the probability of transmission increases if the vehicles are near the VRU cluster. In this paper, we chose broadcast as our transmission mechanism and the VRUs transmitted the packets to all the vehicles within communication range.

For future work, we will try to create a hardware test-bed platform to validate our simulation results. For hardware test-bed we will employ software-defined radios and reconfigure them as UEs and eNodeB.

\bibliographystyle{IEEEtran}
\bibliography{references}

\end{document}